\documentclass{emulateapj}
\usepackage{natbib}

\usepackage{graphicx}
\usepackage{latexsym}
\usepackage{amsfonts,amsmath,amssymb}
\usepackage{url}
\usepackage[utf8]{inputenc}
\usepackage{fancyref}
\usepackage{hyperref}
\hypersetup{colorlinks=false,pdfborder={0 0 0},}

\newcommand{\spin}{a_*}

\newcommand{\Mdot}{\dot{M}}

\newcommand{\Msun}{\rm M_{\sun}}

\newcommand{\fsc}{f_{\rm SC}}

\newcommand{\Rin}{R_{\rm in}}

\newcommand{\Risco}{R_{\rm ISCO}}

\shorttitle{The Spin of LMC X--3}
\shortauthors{Steiner et al.}

\begin{document}

\title{The Low-Spin Black Hole in LMC X--3}

\author{James F.\ Steiner\altaffilmark{1}\altaffilmark{\dag}, 
  Jeffrey E.\ McClintock\altaffilmark{1},   
  Jerome A.\ Orosz\altaffilmark{2}, 
  Ronald A.\ Remillard\altaffilmark{3}, \\
  Charles D.\ Bailyn\altaffilmark{4},
  Mari Kolehmainen\altaffilmark{5},
  Odele Straub\altaffilmark{6} }

\altaffiltext{1}{Harvard-Smithsonian Center for Astrophysics, 60  Garden Street, Cambridge, MA 02138.}
\altaffiltext{2}{Department of Astronomy, San Diego State University, 5500 Campanile Drive, San Diego, CA 92182.}
\altaffiltext{3}{MIT Kavli Institute for Astrophysics and Space  Research, MIT, 70 Vassar Street, Cambridge, MA 02139.}
\altaffiltext{4}{Astronomy Department, Yale University, P.O. Box 208101, New Haven, CT 06520.}
\altaffiltext{5}{Astrophysics, Department of Physics, University of Oxford, Keble Road, Oxford OX1 3RH, UK.}
\altaffiltext{6}{LUTH, Observatoire de Paris, CNRS, Universite Paris Diderot, 5 place Jules Janssen, 92190 Meudon, France.}
\altaffiltext{\dag}{Hubble Fellow.}
\email{jsteiner@cfa.harvard.edu}

\begin{abstract}
  Building upon a new dynamical model for the X-ray binary LMC X--3, we
  measure the spin of its black hole primary via the continuum-fitting
  method. We consider over one thousand thermal-state \emph{RXTE}
  X-ray spectra of LMC X--3. Using a large subset of these spectra, we
  constrain the spin parameter of the black hole to be $\spin =
  0.21^{+0.18}_{-0.22}$ (90\% confidence). Our estimate of the
  uncertainty in $\spin$ takes into account a wide range of systematic
  errors. The low spin and low mass of the black hole 
  further align LMC X--3 with the class of Roche-lobe overflowing,
  transient black-hole systems.  We discuss evidence for a correlation
  between a black hole's spin and the complexity of its X-ray
  spectrum.
\end{abstract}

\keywords{accretion, accretion disks --- black hole physics --- stars:
  individual (\object{LMC X--3}) --- X-rays: binaries}

\section{Introduction} \label{section:intro}

\citet{Leong_1971} discovered LMC X--3 during the first year of the
{\it Uhuru} mission. In 1983, \citet{Cowley_1983} showed via dynamical
observations that the compact X-ray source in this 1.7-day binary is a
black hole.  In \citet{Jerry_LMCX3} we use new optical data to derive
a much-improved dynamical model of the system, which contains a B-type
secondary.  Of chief importance, Orosz et al. report tight constraints
on the orbital inclination angle of the binary, $i = 69.6\pm0.6$~deg,
and the mass of the black hole, $M=6.95\pm0.33~\Msun$.

LMC X--3 is unusual compared to the full assemblage of black-hole
binary systems: On the one hand, like the transient systems, its X-ray
intensity is highly variable because the black hole is fed by
Roche-lobe overflow.  On the other hand, however, the system almost
continually maintains itself in an X-ray bright mode like the
persistent (wind-fed) systems \citep{MNS13, Soria_2001}.

Transient versus persistent black holes are further set apart by the
properties of their black hole primaries: the transient black hole
masses are low and tightly distributed ($7.8~\Msun \pm 1.2~\Msun$),
while the masses of the persistent black holes are appreciably higher
($\gtrsim11~\Msun$; \citealt{Ozel_2010}).  Meanwhile, the spins of the
transients are widely distributed ($\spin \approx 0-1$), while the
three persistent black holes with spin measurements are all high
($\spin \gtrsim 0.85$).  The mode of mass transfer firmly identifies
LMC X--3 as belonging to the class of transient black hole
binaries. The low mass of LMC X--3's primary and its low spin
(reported herein) are perfectly congruent with this classification.

Despite strong variations in X-ray brightness ($>3$ orders of
magnitude), the X-ray spectrum of LMC X--3 is nearly always in a
thermal, disk-dominated state (ideal for measuring spin via the
continuum-fitting method), except during occasional prolonged
excursions into a low-intensity hard state (e.g., \citealt{Smale_2012,
  Wilms_2001}).  Because the X-ray spectrum is strongly
disk-dominated, relatively featureless, and minimally affected by
interstellar absorption, LMC X--3 is a touchstone for testing spectral
models of black-hole accretion disks (e.g.,
\citealt{Kubota_2010,Straub_2011}).  In a precursor to this work, we
fitted all archival X-ray spectra with reliable flux calibration to a
relativistic model of a thin accretion disk.  For hundreds of Rossi
X-ray Timing Explorer ({\it RXTE}) spectra, we showed that the inner
radius of the disk is constant to within $\approx2$\% over a span of
14 years despite the gross variability of the source.  Furthermore,
for an ensemble of eight X-ray missions spanning 26 years, we showed
that the radius is constant to better than $\approx6$\%, despite the
uncertainties associated with cross-calibrating the various detectors.
This result is the strongest observational evidence to date that spin
can be reliably inferred by measuring the inner-disk radius.

The elegant simplicity of a black hole is encapsulated in the famous
"no-hair theorem'', which tells us that an astrophysical black hole is
{\em completely} described by just its mass and spin angular momentum
(the third parameter, electrical charge, being effectively neutral in
astrophysical settings).  A spinning black hole is an enormous
capacitor of angular momentum, with the spin as a ready energy source
that can be tapped mechanically in a black hole's ergosphere.  Spin
has long been proposed as the likely energy source behind the
enormously energetic jets emitted from black holes (e.g.,
\citealt{BZ77}).  This assumption has gained recent support both
theoretically (e.g., \citealt{Tchekhovskoy_2011}), and observationally
(\citealt{NM12,Steiner_jets_2013,MNS13}, but see \citealt{Russell_2013}).

The two primary means by which black hole spins are measured are 1)
X-ray continuum-fitting \citep{Zhang_1997}, and 2) modeling
relativistic reflection (frequently termed the ``Fe-line'' method)
\citep{Fabian_1989}.  The single precept which underpins both methods
is the monotonic relationship between spin and the radius of the
innermost-stable circular orbit (ISCO) for particles orbiting the
black hole.  Consequently, by determining $\Risco$, which is presumed
to be the inner-radius of the accretion disk, one may directly infer a
black hole's angular momentum $J$, usually expressed as the 
dimensionless spin parameter,
\begin{equation}
\spin \equiv cJ/GM^2, \qquad  0 \leq |{\spin}| \leq 1.   
\label{eq:spin}
\end{equation}

The continuum-fitting method -- the basis for the work presented here
-- has been used to estimate the spins of roughly a dozen stellar-mass
black holes (e.g., \citealt{MNS13}, and references therein,
\citealt{Middleton_2006,Mari_GX}).  In this method, the inner-disk
radius is estimated using the thermal, multicolor blackbody continuum
emission from the disk (e.g., \citealt{NT73,SS73}).  The Fe-line
method measures $\Rin$ using the redward extent of relativistic
broadening of reflection features in the disk, and has been applied
just as widely (e.g., \citealt{Brenneman_Reynolds, Walton_2012a,
  Reis_2013a} and \citealt{Reynolds_2013} and references therein).
Both methods, but especially the Fe-line method, are also being
applied to measure the spins of supermassive black holes in active
galactic nuclei (AGN).  LMC X--3 is ideal for continuum-fitting as it
offers a strong, dominantly thermal continuum at nearly all times.
However, its spectra accordingly contain very little signal in
reflection, and so reflection models cannot constrain spin for this
source.

Prior to this investigation, a preliminary estimate of LMC X--3's spin
was obtained via a continuum-fitting measurement from a single {\it
  BeppoSAX} spectrum by \citet{Davis_2006}, later bolstered by
\citet{Kubota_2010}.  Their results were hampered primarily by the
poorly constrained mass available at the time, and so only a rough
estimate was possible, $\spin \sim 0.3$.  Employing the new and
precise mass measurement from \citet{Jerry_LMCX3}, we revisit LMC
X--3's spin determination.  Using the largest data set available to us
-- over a decade of pointed monitoring with ({\it RXTE}) -- we
estimate the spin and its uncertainty for LMC X--3 via X-ray continuum
fitting.

\section{Data and Analysis} \label{section:data}

We replicate the reduction and analysis techniques adopted in
\citet{Steiner_2010} while considering the {\em full} set of {\it
  RXTE} pointed observations over the mission lifetime.  To ensure the
most consistent analysis, we only use data for the PCU-2 detector,
which has the most stable calibration and was most frequently active.
We separate and analyze each segment of continuous exposure, applying
a 300~s lower limit.  For very long observations, we divide the
interval into segments with individual exposure times less than
5000~s.  In total, this yields 1598 spectra with an average exposure
time and net signal of $\sim2$~ks and $\sim 50\times10^3$~counts,
respectively.  The spectra have been background subtracted and
corrected for detector dead time.

We fit the data over the energy range 2.8--25~keV and include a 1\%
systematic error in the count rates in each channel to account for
uncertainties in the response of the detector.  We use the average
PCU-2 spectrum of the Crab to establish our standard absolute flux
calibration \citep{Toor_Seward}\footnote{We have explored the effects
  of the variability of the Crab's spectrum \citep{WilsonHodge_2011}
  on our results and find that they are negligible.}.  For this step,
we use the model {\sc crabcor} \citep{Steiner_2010} which rescales the
normalization by 9.7\% and adjusts the spectral index by
$\Delta\Gamma=0.01$.  There is relatively little interstellar
absorption, $N_{\rm H} = 4\times10^{20}~{\rm cm^{-2}}$
\citep{Page_2003}, which is kept fixed during fitting and modeled via
{\sc tbabs} \citep{tbabs}.

Errors are everywhere quoted at the $1\sigma$ level unless otherwise
indicated.  For the primary -- i.e., thermal disk -- component of
emission, we use {\sc kerrbb2} \citep{MNS13,BHSPEC,KERRBB}, which
incorporates all relativistic effects and directly solves for spin.
The model {\sc kerrbb2} assumes that the disk is razor thin and
optically thick.  To apply {\sc kerrbb2} one must specify four
external input parameters: black hole mass, disk inclination, distance
$D$ and the viscosity parameter $\alpha$ \citep{SS73}.  The Compton
power-law component is modeled empirically using {\sc simpl}
\citep{Steiner_simpl}.  Our complete spectral model is expressed as
{\sc tbabs}({\sc simpl}$\otimes${\sc kerrbb2})$\times${\sc crabcor}.

As an initial step in the analysis, we fit the full data set to our
spectral model using fiducial values for $M$ and $i$
(Section~\ref{section:intro}) and a distance $D = 48.1\pm2.2$~kpc
\citep{Orosz_2009}, and adopting a viscosity $\alpha = 0.03$.  We
include limb darkening and returning radiation effects, adopt zero
torque at the inner-disk boundary, and make the standard assumption
that the black hole's spin axis is aligned with the orbital angular
momentum \citep{Steiner_j1550jets, MNS13}.  There are just four free
fit parameters: $\spin$, mass accretion rate $\Mdot$, photon spectral
index $\Gamma$, and the scattering fraction $\fsc$ (the fraction of
disk photons scattered in the corona).  We constrained the photon
index to lie in the range $\Gamma = 1.5-3.5$; the three other
parameters are unconstrained.

\begin{figure}[tb]
\begin{center}
\includegraphics[width=0.75\columnwidth,angle=90]{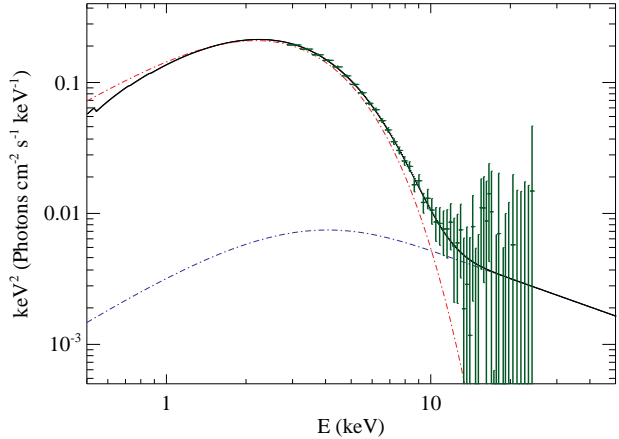}
\caption{A representative fit to a 4~ks exposure of LMC X--3 from 9
  December 1998.  The black line shows the composite model, including
  the effects of photoelectric absorption.  The red and blue
  dash-dotted lines show the intrinsic thermal disk and power-law components,
  respectively. \label{fig:fit} }
\end{center}
\end{figure}

Subsequent to fitting all of the spectra, we use our initial results
to screen out spectra that are unsuitable for the measurement of spin:
We reject data for which $\fsc$ exceeds 25\% (141/1598; see
\citealt{Steiner_2009}).  We likewise discard data for which the
goodness-of-fit $\chi^2_{\nu} > 2$ (4/1598), arriving at a sample of
1454 thermal-state spectra.  Figure~\ref{fig:fit} shows a fit to one
representative spectrum.  The thermal component in red plainly
dominates the flux, and our model of a featureless Comptonized disk
provides a more-than sufficient fit ($\chi^2_{\nu} \approx 0.5$), with
only four free parameters.

A restriction of {\sc kerrbb2} is that it is only applicable to those
data firmly in the ``thin-disk'' limit (scale height $H/R << 1$).  The
scale-height of the X-ray emitting region of the disk is determined by
radiation pressure, itself set by the luminosity.  Previous work has
demonstrated that the presence of a geometrically-thin,
optically-thick disk can be reliably assumed over the luminosity range
$L \approx 5-30\% L_{\rm Edd}$, where the Eddington-luminosity $L_{\rm
  Edd} \approx 1.3 \times 10^{38} ( M/\Msun )$~erg~s$^{-1}$
\citep{MNS13}.

\section{Results}\label{section:results}

In our initial run, a total of 391 spectra fulfilling the thermal
selection also match this luminosity (thin-disk) criterion.  Using the
spectra which make both cuts, we obtain a spin  $\spin = 0.190 \pm
0.005$ (weighted mean and its 1$\sigma$ error).

A comprehensive run is now performed to assess the error in spin from
uncertainties in the other measurement quantities.  For each point
along a grid of $M$, $i$, and $D$, we repeat our spectral fits to the
pre-selection of 1454 spectra. Mass is sampled from $M: 5-11~\Msun$,
inclination $i: 60\degr-75\degr$, and $D: 41-56$~{\rm kpc}.  The
values of $L/L_{\rm Edd}$ depend upon $M,i,$ and $D$, but typically
$\sim 400$ of these spectra fulfill the luminosity criterion.  At {\em
  each} gridpoint, the distribution of spin from the selected spectra
gives a single, mean spin, and its uncertainty\footnote{As demonstrated in
\citet{Steiner_2010}, the inner disk radius (which corresponds to a
particular value of spin) at any grid point has a spread of $<5\%$,
while the mean of the distribution is determined with much greater
precision.}.  By applying weights according to the probability of
each gridpoint (calculated using the measurements of $M$, $i$, and $D$
given in Sections~\ref{section:intro},\ref{section:data}), results are
combined to achieve a composite distribution in spin.  This
distribution is inclusive of {\em all measurement errors}.

In previous work, we have assessed the effects of a wide range of
systematic errors \citep{Steiner_2010,Steiner_j1550spin_2011} and
found that only two are significant: the uncertainty in $\alpha$ and
the uncertainty in the spectrum of the Crab, which we use as our flux
standard (Section~\ref{section:data}).  We take uncertainty in
$\alpha$ into account by repeating the analysis for $\alpha = 0.01$
and $\alpha = 0.1$ and then averaging the spin distributions,
weighting them equally.  We incorporate a 10\% uncertainty in the
absolute X-ray flux calibration \citep{Toor_Seward} by broadening our
distribution in $\Rin$ using a Gaussian kernel with 5\% width.
Finally, an additional 2\% broadening is used to account for the small
variation introduced by adopting a different choice of Comptonization
model (\citealt{Steiner_2010}).

\begin{figure}[tb]
\begin{center}
\includegraphics[width=0.75\columnwidth,angle=90]{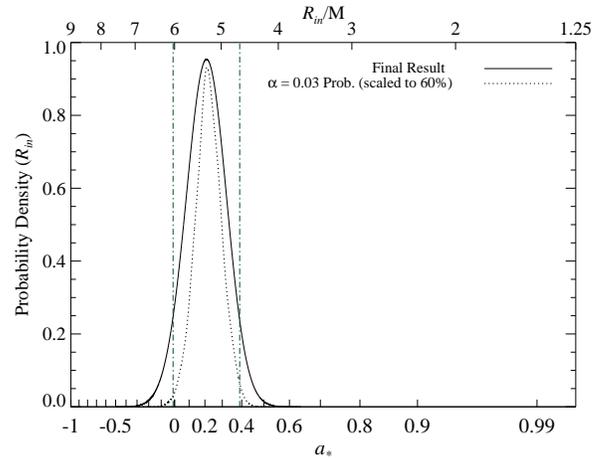}
\caption{The net probability distribution of $\spin$ ($R_{\rm in}$).
  The abcissa is scaled logarithmically in $\Rin$, a metric which
  reflects our measurement scale.  The corresponding quantity of chief
  physical interest, namely $a_*$, is shown on the lower axis.  Our
  final distribution (solid line) incorporates measurement error and
  systematic error, most notably uncertainties in the
  $\alpha$-viscosity and the absolute flux calibration.  The pair of
  dash-dotted lines indicate the 90\% confidence interval.  The dotted
  line shows our result for $\alpha = 0.03$ ignoring the effects of
  systematic error (where the probability distribution has been
  rescaled for comparison).
\label{fig:spindist}}
\end{center}
\end{figure}

The combined result is shown in Figure~\ref{fig:spindist}.  Our {\em
  final} measurement of spin, including all measurement {\em and}
systematic uncertainties is $\spin = 0.21^{+0.18}_{-0.22}$ (90\%).
For comparison, our measurement without considering systematic errors
(using $\alpha = 0.03$) yielded $\spin = 0.21 \pm 0.10$ (90\%).
Table~\ref{tab:results} gives our final determination of spin as
measured at several confidence levels, explicitly given because the
measurement errors in spin are generally nonlinear.

As a bottom line, LMC X--3 has a precisely determined spin, which is low.

  \begin{deluxetable}{lcc}
  \tabletypesize{\scriptsize}
  \tablecolumns{4}
  \tablewidth{0pc}
  \tablecaption{Final Spin Determination}
  \tablehead{    \colhead{Confidence Level} & \colhead{Spin Interval} & \colhead{$R_{\rm in}$ Interval} \\}
 \startdata
 68\% ($1\sigma$)        & $\spin = 0.21\pm0.12 $              & $\Rin/M = 5.3\pm0.4  $  \\
90\%                    &  $ 0.21^{+0.18}_{-0.22} $     & $ 5.3^{+0.7}_{-0.6}  $  \\
95\% ($2\sigma$)        &  $ 0.21^{+0.21}_{-0.27} $     & $ 5.3^{+0.9}_{-0.8}  $  \\
99.7\% ($3\sigma$)      &  $ 0.21^{+0.30}_{-0.44} $     & $5.3^{+1.4}_{-1.1}  $  
\enddata
\tablecomments{Incremental confidence intervals for our final,
    adopted spin result, which accounts for all sources of error.}
\label{tab:results}
\end{deluxetable}

\section{Discussion}\label{section:discussion}

In Section~\ref{section:results} we considered the effects of all
measurement errors and known systematic errors on our estimate of
$a_*$ (apart from the assumption that the black hole's spin is aligned
with the orbital angular momentum vector; see Section 5.4 in
\citealt{MNS13}).  In this section we consider the error we incur by our
reliance on the Novikov-Thorne model and conclude that it is
relatively small.  We then compare the low spin of LMC X--3 to the
spins of several other sources, tentatively concluding that the
Compton component is relatively weak for low spin black holes and
strong for rapidly spinning ones.

\subsection{Errors from the Novikov-Thorne Model}

While our earlier study of LMC X--3 \citep{Steiner_2010}, and similar
studies of other black hole binaries (e.g., \citealt{Gierlinski_2004}),
provide compelling evidence that the inner disk radius is constant in
black hole binaries, it does not prove that this radius is in fact
$R_{\rm ISCO}$.  The central assumption of the thin-disk model is that
the viscous torque vanishes at the ISCO and that no flux is emitted
from within this radius.  This assumption has been tested by several
groups using sophisticated general-relativistic magnetohydrodynamic
(GRMHD) codes.  The results from two groups are exemplified by three
key studies, each aimed at testing the reliability of spin estimates.
Results from one group are reported in \citet{Noble_2011}, and from
the other in \citet{Kulkarni_2011} and \citet{Zhu_2012}.  Both groups
produce synthetic observations of their simulations as they would
appear to a distant observer at a range of viewing angles.  

Although there are subtle differences in the approaches taken by the
two groups, one can make a reasonably direct comparison between
\citet{Kulkarni_2011} and \citet{Noble_2011}.  Both groups used ad hoc
but reasonable cooling prescriptions in post-processing to convert
magnetic stresses into radiation.  By treating the dissipation as
local and thermal, disk spectra have been generated for a nonspinning
black hole by Noble et al. and for black holes with a range of spins
by Kulkarni et al.  From these two works, one concludes that spin
measurements systematically {\em overestimate} the spin, that this
effect is most pronounced at high inclination, and that the fractional
change in $\Rin$ is independent of $\spin$.  The deviation of
$\Rin/\Risco$ from unity is of order ten percent.  However, the
state-of-the-art analysis has been achieved by \citet{Zhu_2012}, and
its findings differ appreciably, as we now discuss.

Zhu et al.\ include, as an additional post-processing step, full
radiative-transfer through the disk atmosphere in generating the
simulated GRMHD spectra.  In contrast to the $\sim10$\% shift in
$\Rin/\Risco$ reported by \citet{Kulkarni_2011} and
\citet{Noble_2011}, Zhu et al. find a much smaller deviation because
their more sophisticated approach identifies a hard power-law
component of emission that is largely produced inside $\Risco$.  It is
this component, which in the earlier work was lumped in with the
thermal emission, that was largely responsible for the shift in
$\Rin/\Risco$.  Analyzing their simulated spectra using the model in
Section~\ref{section:data}, Zhu et al. find that the shift in
$\Rin/\Risco$ is only $\sim 3\%\pm2\%$ and that it depends only weakly
on inclination, $\alpha$, and luminosity (see Table~2 in
\citealt{Zhu_2012}).

In short, deviations from Novikov-Thorne are likely of minor
consequence.  For the nominal spin of LMC X--3, a 3\% offset in $\Rin$
would imply $\spin \approx 0.16$ (a shift of ${\Delta}\spin=-0.05$),
or a 0.4$\sigma$ correction to our final result.

\subsection{A Possible Link Between Spin and Spectral Complexity}

Although the spins of more than a dozen stellar-mass black holes have
been measured, only two have spins as low as LMC X--3, namely,
A0620--00  and H1743-322 \citep{MNS13}.  The spectra of
A0620--00 and LMC X--3 (at luminosities $>0.05L_{\rm Edd}$) are
remarkably simple, consisting of a dominant thermal component with
Compton power-law and reflection components that are always quite
faint.  The spectrum of H1743--322 is not as consistently simple.
However, for a large sample of 65 thermal spectra, the Compton (and
reflection) component of this source is very faint with a mean
strength of $f_{\rm SC}$ = 1.2\% \citep{Steiner_2009}, weaker than for
any other sources with a spin measurement (apart from A0260--00 and
LMC X--3).

The simplicity of the spectra of these three sources contrasts sharply
with the spectra of the two extreme-spin sources, Cyg X--1 and GRS 1915+105 
(\citealt{MNS13}, and references therein), and to a lesser degree the spectrum
of LMC X--1 ($a_* = 0.92_{-0.07}^{+0.05}$; \citealt{Gou_2009}).  The
spectra of these sources are generally complex; i.e., they show strong
rms variability (from their power-density spectra), are strongly
Comptonized, and contain prominent reflection features.  Comparison
with the simple spectra of A0620--00, LMC X--3 and H1743--322 suggests
that spectral complexity is correlated with spin.

A prediction of this hypothesis is that the spin of GS~2000+25, whose
spectral properties (and outburst light curve) are very similar to
that of A0620--00 \citep{Terada_2002}, is very low, and that the spin
of V404 Cyg, whose spectrum was consistently observed to be strongly
Comptonized \citep{Tanaka_Lewin}, is quite high.  One other source may
be able to provide a strong test of our prediction: 4U1957+11.
Although the spin and mass are presently unknown, there are
indications that the former is likely high and the latter likely
low \citep{Nowak_2012}, while its spectrum is very thermal and
generally only weakly Comptonized.  Confirmation of an extreme spin
for 4U 1957+11 could decisively rule out this hypothesis.

\section{Conclusions}\label{section:concs}

We have analyzed all 1598 spectra of LMC X--3 collected during the
{\it RXTE} mission.  Using a selected sample of $\approx400$ spectra,
our precise measurement of black hole mass and inclination \citep{Jerry_LMCX3}, and
the continuum-fitting method, we derive a strong constraint on the
spin of the black hole: $\spin = 0.21^{+0.18}_{-0.22}$ (90\%
confidence).  Our comprehensive error estimate takes into account all
known sources of uncertainty, e.g., the uncertainties in $M$, $i$,
$D$, $\alpha$, and in the absolute X-ray flux calibration.

The simple and predominately thermal spectra of LMC X--3 and
A0620--00, the black holes with the lowest measured spins, contrast
sharply with the complex and strongly Comptonized spectra of
GRS~1915+105 and Cyg X--1, the two black holes with near-extreme spin.
This comparison suggests a possible link between spin and the degree
of spectral complexity, a hypothesis that can be tested, and which
predicts a low spin for GS~2000+25 and a high spin for V404 Cyg.  The
black hole 4U 1957+11 may allow a falsification of the hypothesis, if
its spin can be measured.

By virtue of the no-hair theorem, we have a complete and quite precise
description of the black hole in LMC X--3.  These three -- and only three --
characteristics define the black hole in entirety: a charge of zero, $M \approx
7.0~\Msun$ and $\spin \approx 0.2$.

\acknowledgments

This work was made possible by the Odyssey computing cluster,
supported by the FAS Science Division Research Computing Group at
Harvard University.  We thank Chris Done for enlivening discussions on
LMC X--3 and Colleen Hodge-Wilson for providing data on the Crab's
variability.  Support for JEM has been provided by NASA grant
NNX11AD08G, and for JFS by NASA Hubble Fellowship grant
HST-HF-51315.01. \\

{\it Facility:} \facility{RXTE}

\end{document}